\documentclass[aps,prl,twocolumn,superscriptaddress,showpacs,preprintnumbers]{revtex4} 

\usepackage{setspace}
\usepackage{graphicx}  
\usepackage{dcolumn}   
\usepackage{bm}        
\usepackage{rotating}
\usepackage{hyperref}

\usepackage{preprintcover}  
\PreprintCoverPaperTitle{Search for Dark Matter Candidates and Large Extra Dimensions in events with a 
photon and missing transverse momentum in $pp$ collision data at $\sqrt{s}=$7~TeV with the ATLAS detector}  
\PreprintIdNumber{CERN-PH-EP-2012-209}  
\PreprintCoverAbstract{Results of a search for new phenomena in events with an energetic photon
and large missing transverse momentum in proton-proton collisions at $\sqrt{s}=7$~TeV
are reported.
Data collected by the ATLAS experiment at the LHC
 corresponding to an integrated luminosity of 4.6~fb${}^{-1}$ are used.
Good agreement is observed between the data and the Standard Model predictions.
The results are translated into exclusion limits on models with
large extra spatial dimensions and on pair production of weakly interacting
dark matter candidates.
}  
\PreprintJournalName{Physical Review Letters}  

\newcommand{\met}{E_{\rm T}^{\rm miss}}

\newcommand{\akt}{\hbox{anti-${k_t}$} }
\newcommand{\pt}{p_{\rm T}}

\newcommand{\zll}{Z/\gamma^*(\rightarrow \ell^+ \ell^-)}

\newcommand{\znn}{Z(\rightarrow \nu \bar{\nu})}

\newcommand{\wln}{W(\rightarrow \ell \nu)}
\newcommand{\mchi}{m_{\chi}}

\begin{document}


\title{Search for Dark Matter Candidates and Large Extra Dimensions in events with a  
photon and missing transverse momentum in $pp$ collision data at $\sqrt{s}=$7~TeV with the ATLAS detector}

\author{The ATLAS Collaboration}


\begin{abstract}
Results of a search for new phenomena in events with an energetic photon 
and large missing transverse momentum in proton-proton collisions at $\sqrt{s}=7$~TeV
are reported. 
Data collected by the ATLAS experiment at the LHC   
 corresponding to an integrated luminosity of 4.6~fb${}^{-1}$ are used.
Good agreement is observed between the data and the Standard Model predictions.
The results are translated into exclusion limits on models with
large extra spatial dimensions and on pair production of weakly interacting 
dark matter candidates.
\end{abstract}

\maketitle

\hyphenation{ATLAS}


Events with an energetic  photon and 
large missing momentum   in the final state 
constitute a clean and distinctive signature in searches for new physics at colliders. 
In particular,  monophoton and monojet final states have been
studied~\cite{lep,d0,cdf_add,cdf_wimps,cmsmonojet, cms, atlasmonojet,atlas_monojet_new} in the context of searches 
for supersymmetry and large extra spatial dimensions (LED), aiming to provide a solution to the
mass hierarchy problem, and the search for weakly interacting massive particles (WIMPs) as candidates for 
dark matter (DM).

The Arkani-Hamed, Dimopoulos, and Dvali (ADD) model for LED~\cite{add}     
explains the large difference between the electroweak unification scale $O(10^2$)~GeV and 
the Planck scale   $M_{Pl} \sim O(10^{19}$)~GeV     
by postulating the presence of $n$ extra spatial dimensions of size $R$,  
and defining a fundamental Planck scale  in $4+n$ dimensions, $M_D$, given by 
${M_{Pl}}^2 \sim {M_D}^{2+n}R^n$. 
The extra spatial dimensions are compactified,
 resulting in a Kaluza-Klein tower of massive graviton modes.
At hadron colliders, these graviton modes may escape detection and 
can be produced in association with an energetic
photon or a jet, leading to a monophoton or monojet signature.

The presence of a non-baryonic DM component in the universe is inferred 
from the observation of its gravitational interactions~\cite{review}, although   
its nature is otherwise unknown. A WIMP $\chi$ 
with mass $\mchi$ in the range between 1~GeV  and a few TeV is  
a  plausible candidate for DM.  It could be detected via its 
scattering with heavy nuclei~\cite{direct}, the detection of cosmic rays (energetic photons, electrons, positrons, protons, antiprotons, or neutrinos) from $\chi \bar{\chi}$ annihilation in astrophysical sources~\cite{review}, or via $\chi \bar{\chi}$ pair-production at colliders  
where the WIMPs do not interact with the detector and the event is identified 
by the presence of an energetic photon or jet from initial-state radiation. 
The interaction of  WIMPs with Standard Model (SM) particles 
is assumed to be driven by a mediator with mass at the TeV scale and 
described using a non-renormalizable effective theory~\cite{wimps} 
with several operators. The vertex coupling is suppressed by an effective cut-off mass 
scale $M_* \sim M/\sqrt{g_1 g_2}$, 
where $M$ denotes the mass of the mediator and $g_1$ and $g_2$ are 
the couplings of the mediator to the WIMP and SM particles.

This Letter reports results of the search for new phenomena in monophoton final states, 
based on $\sqrt{s}=7$~TeV
proton-proton collision data
corresponding to an integrated luminosity of 
4.6~fb${}^{-1}$  collected with the ATLAS detector at the LHC during 2011. 
The ATLAS detector is described in detail elsewhere~\cite{atlas}. 
The data are collected using a three-level trigger system that selects events with  
missing transverse momentum greater than $70$~GeV.
In the analysis, events are required to have $\met > 150$~GeV, where 
$\met$ is computed as the magnitude of the vector sum of the transverse momentum of all 
noise-suppressed calorimeter topological clusters with $|\eta| < 4.9$~\cite{met,coord}.
A photon is also required with transverse momentum  
$\pt > 150$~GeV and $|\eta| < 2.37$, excluding the calorimeter 
barrel/end-cap transition regions $1.37<|\eta|<1.52$~\cite{atlas}.
With these criteria,  the trigger selection is more than 98$\%$  efficient, as determined using events selected with a muon trigger. 
The cluster energies 
are corrected 
for 
the different response of the calorimeters 
to hadronic jets, $\tau$ leptons, electrons or photons, 
as well as dead material and out-of-cluster energy losses.
The photon candidate must pass {\it tight}
identification criteria~\cite{photon} 
and 
is required to be isolated:  
the energy not associated 
with the photon cluster in a cone of radius $\Delta R=\sqrt{(\Delta \eta)^2 + (\Delta \phi)^2}=0.4$ around 
the candidate is required to be less than 5~GeV.
Jets are defined using the $\akt$ jet algorithm~\cite{akt} with the distance parameter 
set to $R=0.4$. 
The measured 
jet $\pt$ is corrected for detector effects
and for
contributions from multiple proton-proton interactions per beam bunch crossing (pileup)~\cite{jetcor}.  

Events with more than one jet with $\pt > 30$~GeV and $|\eta|< 4.5$ are rejected. 
Events with one jet are retained to increase the signal acceptance and reduce 
systematic uncertainties related to the modeling of initial-state radiation.
The   
reconstructed photon, $\met$ vector and jets (if found) are required to be well separated in the transverse plane with $\Delta \phi(\gamma, \met) > 0.4$, $\Delta R(\gamma,\rm{jet}) >0.4$, and $\Delta \phi({\rm{jet}},\met) > 0.4$. 
Additional quality criteria~\cite{jetq} are  applied to ensure that jets and photons 
are not produced by noisy calorimeter cells, and to avoid problematic detector regions. 
Events with identified electrons or muons are vetoed to reject mainly $W/Z$+jets and
$W/Z+\gamma$ background processes with charged leptons in the final state. Electron (muon) candidates are required to have $\pt > 20$~GeV and $|\eta| <
2.47$ ($\pt > 10$~GeV and $|\eta| <
2.4$), and to pass the {\it medium} {\it{(combined})} criteria~\cite{emu}.
The final data sample contains 116 events, where 88 and 28 events have 
zero and one jet, respectively.

The SM background to the monophoton signal is dominated by the 
irreducible $Z(\to \nu\bar{\nu})+\gamma$ process, and receives contributions from 
$W/Z+\gamma$ events with unidentified electrons, muons or hadronic $\tau$ decays, and $W/Z$+jets events 
with 
an electron or jet misreconstructed as a photon. In addition, the monophoton sample
receives small contributions from top-quark, $\gamma \gamma$, diboson ($WW,ZZ,WZ$), $\gamma$+jets, and multi-jet processes. 

Background samples of simulated
$W/Z+\gamma$ events are generated using {\sc alpgen}~\cite{alpgen}, interfaced to 
{\sc herwig}~\cite{herwig} with {\sc jimmy}~\cite{jimmy}, and {\sc Sherpa}~\cite{sherpa}, using  CTEQ6L1~\cite{cteq} parton distribution functions (PDFs) and requiring a minimum 
photon $\pt$ of 40~GeV.
  Background samples of $W/Z$+jets and $\gamma$+jets 
processes are generated  
using {\sc alpgen} plus {\sc herwig}/{\sc jimmy}, with  CTEQ6L1 PDFs.
Top-quark production samples are generated  
using {\sc mc@nlo}~\cite{mcatnlo} and CT10~\cite{ct10} PDFs, while diboson processes are generated
using {\sc herwig}/{\sc jimmy} normalized to next-to-leading-order (NLO) predictions with MRST2007~\cite{mrst} PDFs. 
Multi-jet and 
$\gamma \gamma$ processes  
are generated using {\sc pythia~6}~\cite{pythia} with MRST2007 PDFs.   

Signal Monte Carlo (MC) samples are generated according to the 
ADD model using 
the {\sc pythia~8} leading-order (LO) 
perturbative QCD (pQCD)  
implementation with default settings, requiring a minimum photon  $\pt$ of 80~GeV, and
an ATLAS tune for the underlying event (UE) contribution~\cite{atlastune} 
including the CTEQ6L1 PDFs. The number of extra dimensions $n$ is varied from 2 to 6 and values of $M_D$ in the $1-2$~TeV range are considered.  
For consistency with a previous monojet analysis performed 
in ATLAS~\cite{atlasmonojet,atlas_monojet_new}, the yields corresponding to CTEQ6.6~\cite{cteq66} PDFs are 
used,  as obtained by reweighting these samples.
The samples are normalized to NLO predictions~\cite{nlo}. The  LO-to-NLO normalization 
factors decrease from 1.5 to 1.1  as  $n$  increases.

Simulated events corresponding to the $\chi \bar{\chi} + \gamma$ process with a minimum photon $\pt$ of 80~GeV
are generated using LO matrix elements from  {\sc madgraph}~\cite{madgraph}  interfaced to {\sc pythia~6}  
using CTEQ6L1 PDFs. 
Values for $\mchi$ between 1~GeV and 1.3~TeV are considered.
In this analysis, WIMPs are assumed to be Dirac fermions and the vertex operator is taken to have the structure of a scalar, vector, axial-vector or tensor, corresponding respectively to the operators D1, D5, D8 and D9 in Refs.~\cite{wimps,masses}. 
These operators correspond to 
spin-independent (D1 and D5) and spin-dependent (D8 and D9) interactions.
The MC samples are  passed through a full simulation~\cite{atlas_sim} 
of the ATLAS detector and trigger system, based on GEANT4~\cite{geant}. The simulated events are
reconstructed and analyzed with the same analysis chain as the data.

The normalization of the MC 
predictions for the dominant $W/Z+\gamma$  background processes are set using 
scale factors  determined in a data 
control sample, resulting in a significant reduction of the background uncertainties.    
A $\gamma+\mu+\met$ control sample with an identified muon  
is defined by inverting the muon veto in the nominal event selection criteria discussed above. 
According to the simulation, the sample contains a 
71$\%$ (19$\%$) contribution from $W+\gamma$ ($Z+\gamma$) processes.  
This control sample is used to normalize separately the $W+\gamma$ and $Z+\gamma$ MC predictions   
determined by {\sc alpgen} and {\sc Sherpa}, respectively. 
In each case, the scale factor is defined as the ratio of the data to the given MC prediction,  
after the contributions from the rest of the  
background processes are subtracted. 
The scale factors, 
extracted simultaneously
to take into account correlations, are    
$k (W+\gamma)= 1.0\pm0.2 $ and  
$k (Z+\gamma)= 1.1\pm0.2 $, where statistical and systematic uncertainties are included. 

Dedicated studies are performed to determine  the probability for electrons or jets to be identified as photons, 
resulting in data-driven estimates of $W/Z$+jet background contributions.
A data sample of $Z$ boson candidates is employed  to compute   
the fraction of electrons from the $Z$ boson decay that are
reconstructed as photons. This fraction decreases from 2$\%$ to 1$\%$
as $\pt$ increases from 150~GeV to 300~GeV, and increases from 1$\%$ to $3\%$ as 
$|\eta|$ increases.
 These rates are employed to determine the $W(\to e \nu)$+jets
background in the signal region, for which a control data sample selected with  the nominal selection 
criteria and an electron instead of a photon is used. This results in a 
total $W(\to e \nu)$+jet background estimation of $14\pm6$ events, where the uncertainty  
is dominated by the limited size of the control data sample.  
Control samples enhanced in jets identified as photons are defined using nominal selection criteria with 
non-isolated photon candidates and/or  photon candidates 
passing a {\it loose} selection~\cite{photon} but not the nominal 
identification requirements. 
The ratio of isolated to non-isolated photons in the 
loose-photon selected sample together with the number of non-isolated photons passing the nominal identification requirements are used to determine the rate of jets identified as photons in the signal region, after the contribution from $W/Z+\gamma$ processes has been subtracted. This gives an estimate of 
$4.3\pm1.9$ $W/Z$+jet background events. 

The $\gamma$+jet and multi-jet background contributions to the signature of a photon 
and large $\met$ originate from the misreconstruction of the energy of a
jet in the calorimeter. The direction of $\met$ vector therefore tends to be aligned with the 
jet.  These background contributions  are determined from data using a control sample
with the nominal selection criteria and at least one jet with $\pt >30$~GeV and $\Delta\phi({\rm{jet}},\met) < 0.4$.
After the subtraction of electroweak boson and top-quark production processes, 
a linear 
extrapolation of the measured $\pt$ spectrum  to $\pt < 30$~GeV leads to 
an estimate of $1.0\pm0.5$ background events in the signal region, where the uncertainty
is due to the ambiguity in the functional form used in the extrapolation. 
Background contributions from top-quark, $\gamma \gamma$, and diboson production processes,  
determined using MC samples, are small. Finally, non-collision backgrounds are negligible.

A detailed study of systematic uncertainties on the background predictions has been performed. 
An uncertainty of 0.3$\%$ to $1.5\%$  on the absolute photon energy scale~\cite{photon}, depending on the 
photon $\pt$ and $\eta$, translates into a $0.9 \%$ uncertainty on the total background prediction.
Uncertainties on the simulated photon energy resolution, photon isolation, and photon identification 
efficiency introduce a combined 1.1$\%$ uncertainty on the background yield.
Uncertainties on the simulated lepton identification efficiencies introduce a $0.3\%$ uncertainty on the 
background predictions.
The uncertainty on the absolute jet energy scale~\cite{jetcor} and jet energy resolution 
introduce $0.9\%$ and $1.2\%$ 
uncertainties on the background estimation, respectively.
A 10$\%$ uncertainty on the absolute energy scale for low $\pt$ jets and 
unclustered energy in the calorimeter, and a $6.6\%$ uncertainty on the subtraction of 
pileup contributions, are taken into account. They affect the $\met$ determination  and translate 
into $0.8\%$ and $0.3\%$ uncertainties 
on the background yield, respectively.  
 The dependence of the predicted 
$W/Z+\gamma$  backgrounds on 
the parton shower and hadronization model used in the MC simulations 
is studied by comparing the predictions from {\sc Sherpa} and  {\sc alpgen}. 
This results in a conservative 6.9$\%$  uncertainty on the total background yield. 
Uncertainties due to the choice of 
PDFs and the variation of the renormalization and factorization scales in the $W/Z+\gamma$ 
MC samples introduce an additional 1.0$\%$ uncertainty on the total background yields.  
Other sources of systematic uncertainty related to the trigger 
selection, the lepton $\pt$ scale and resolution, the pileup description,  
background normalization of the top quark, $\gamma \gamma$ and diboson contributions,  
and a $1.8\%$ uncertainty on the total luminosity~\cite{lumi} 
introduce a combined uncertainty of  less than 0.5$\%$ on the total predicted yields.  
%
 The different sources of uncertainty are added in
quadrature, resulting in a total 15$\%$ uncertainty on the background prediction.

\begin{figure}[h]
\begin{center}
\includegraphics[width=0.45\textwidth]{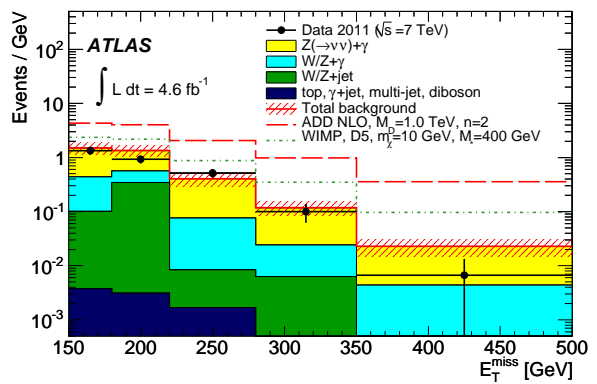}
\end{center}
\caption{\small
The measured $\met$ distribution (black dots) compared to
the SM (solid lines), SM+ADD  (dashed lines),  and SM+WIMP
(dotted lines) predictions, for two particular ADD  and WIMP scenarios. 
}
\label{fig:ptmet}
\end{figure}

In Table~\ref{tab:results}, the observed number of events and the SM predictions 
are presented. 
The data are in agreement with the SM background-only hypothesis with 
a $p$-value of 0.2.
Figure~\ref{fig:ptmet} shows the measured $\met$ distribution compared to the
background predictions. 
The results are expressed in terms of  model-independent 90$\%$ and 95$\%$ confidence level (CL) upper limits 
on the visible cross section, defined as the production cross section times acceptance times efficiency ($\sigma \times A \times \epsilon$),   
using the $CL_s$ modified frequentist approach~\cite{cls} and 
considering  the systematic 
uncertainties on the SM backgrounds and on the integrated luminosity. Values of $\sigma \times A \times \epsilon$ 
above 5.6~fb and 6.8~fb are excluded at 90$\%$ CL and 95$\%$ CL, respectively.
Typical event selection efficiencies of 
$\epsilon \sim 75\%$ are found in simulated ADD and WIMP signal samples.   


\begin{table}[h]
\begin{center}
\begin{footnotesize}
\renewcommand{\baselinestretch}{1.2}
\begin{tabular}{l l l l} \hline\hline
Background source & Prediction & $\pm$ (stat.) & $\pm$ (syst.) \\ \hline
$\znn + \gamma$     &  $93$  & $\pm$ 16  &   $\pm$ 8  \\ 
$\zll + \gamma$     &  $0.4$ & $\pm$ 0.2 & $\pm$ 0.1\\
$\wln + \gamma$     &  $24$  & $\pm$ 5   & $\pm$ 2\\
$W/Z$ + jets        &  $18$  &  $-$        & $\pm$ 6\\
Top                 &  $0.07$& $\pm$ 0.07& $\pm$ 0.01  \\
$WW,WZ,ZZ, \gamma \gamma$      & $0.3$&$\pm$ 0.1&$\pm$ 0.1 \\   
$\gamma$+jets and multi-jet & $1.0$&$-$&$\pm$ 0.5\\
Total background         & $137$&$\pm$ 18&$\pm$ 9\\ \hline
Events in data  (4.6 fb${}^{-1}$)& 116 \\ \hline\hline
 \end{tabular}
\end{footnotesize}
\end{center}
\renewcommand{\baselinestretch}{1}
\caption{
\small
The number of events in data compared to the SM predictions, including statistical and systematic uncertainties.
In the case of  $W/Z$ + jets, $\gamma$+jets and multi-jet processes a global uncertainty is quoted.
\label{tab:results}
}
\end{table}

\begin{figure}[h]
\begin{center}
\includegraphics[width=0.45\textwidth]{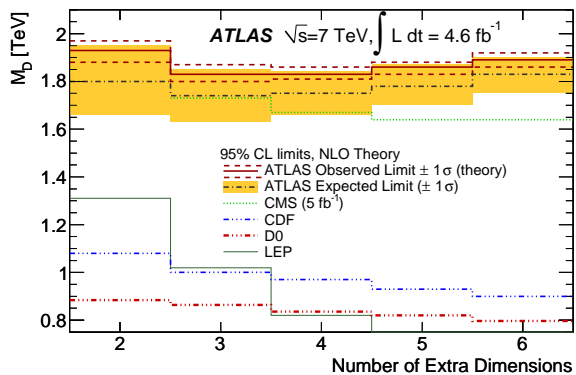}
\end{center}
\caption{\small
Observed (solid lines) and expected (dashed-dotted lines)  95$\%$ CL limits on $M_D$ as a function of the 
number of extra spatial dimensions $n$ in the ADD  model. 
The results are compared with previous results~\cite{lep,d0,cdf_add,cms} (other lines). 
}
\label{fig:add}
\end{figure}

The results are translated into $95\%$~CL limits  on the parameters of the ADD model.
The typical $A \times \epsilon$ of the selection criteria is $20.0\pm0.4 {\rm (stat.)}\pm1.6 {\rm  (syst.)} \%$,  approximately 
independent of $n$ and $M_D$. Experimental uncertainties related to the photon, jet and $\met$
scales and resolutions, the photon reconstruction, the trigger efficiency, the pileup description, and the luminosity 
introduce a 6.8$\%$ uncertainty on the 
signal yield. 
Uncertainties related to the modeling of the initial- and final-state 
gluon radiation 
translate into a 3.5$\%$ uncertainty on the ADD 
signal yield. 
Systematic uncertainties due to PDFs   
result in a 0.8$\%$ to 1.4$\%$ uncertainty on the signal $A \times \epsilon$  
and a $4\%$ to 11$\%$ uncertainty on the signal cross section, increasing as $n$ increases.
Variations of the renormalization and factorization scales by factors of two and one-half
introduce a 0.6$\%$ uncertainty on the signal $A \times \epsilon$ and an uncertainty on the  
signal cross section that decreases from 9$\%$ to 5$\%$ as $n$ increases.
Figure~\ref{fig:add} shows the expected and observed 95$\%$~CL lower limits on $M_D$ as a function of $n$, as
determined using the $CL_s$ method and considering uncertainties on both signal and SM background predictions.
Values of $M_D$ below 1.93~TeV ($n=2$), 1.83~TeV ($n=3$ or $4$), 1.86~TeV ($n=5$), and 1.89~TeV ($n=6$) are excluded at 95$\%$~CL. The observed limits decrease by $3\%$ to $2\%$  after 
considering the $-1\sigma$ uncertainty from PDFs, scale variations, and parton shower modeling 
in the ADD theoretical predictions (dashed lines in Figure~\ref{fig:add}).  
These results improve upon previous limits on $M_D$ from LEP and Tevatron experiments~\cite{lep,d0,cdf_add}. 
In this analysis, no weights are applied for signal events in the phase space region with $\hat{s} > M_D^2$,  which is
sensitive to the unknown ultraviolet behavior of the theory. 
For $M_D$ values 
close to the observed limits, 
the  visible signal cross sections decrease by $15\%$ to $75\%$ as $n$ increases when  
truncated samples with 
$\hat{s} < M_D^2$ are considered. 
This analysis 
probes a kinematic range for which the model predictions are
defined but ambiguous.

\begin{figure}[h]
\begin{center}
\includegraphics[width=0.45\textwidth]{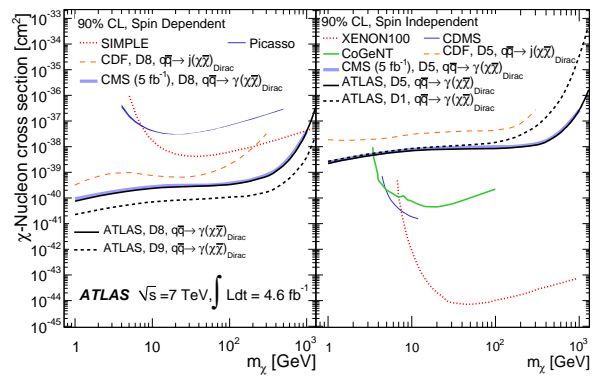}
\end{center}
\caption{\small
90$\%$ CL upper limits on the nucleon-WIMP
cross section as a function of $\mchi$ for spin-dependent (left) and spin-independent (right) 
interactions~\cite{wimps,fixed}.
The results are compared with previous results~\cite{cdf_wimps,cms,direct}.
%
}
\label{fig:wimps}
\end{figure}

Similarly, 90$\%$~CL upper limits on the pair production cross section of dark matter WIMP candidates are determined.
The $A \times \epsilon$ of the selection criteria 
are typically $11.0\pm0.2 {\rm (stat.)}\pm1.6 {\rm (syst.)}\%$ for the D1 operator, $18.0\pm0.3 {\rm (stat.)}\pm1.4 {\rm (syst.)}\%$ for the D5 and D8 operators, and $23.0\pm0.3 {\rm (stat.)}\pm2.1 {\rm (syst.)}\%$ for the D9 operator, 
with a moderate dependence on $\mchi$.
Experimental uncertainties, as discussed above, 
translate into a 6.6$\%$ uncertainty on the signal yields. 
Theoretical uncertainties on initial- and final-state gluon radiation introduce 
a $3.5 \%$ to $10\%$ uncertainty on the signal yields.  
The uncertainties related to PDFs result in $1.0\%$ to $8.0\%$  and $5.0\%$ to $30\%$ 
uncertainties on the signal $A \times \epsilon$ and cross section, respectively.
Variations of the renormalization and factorization scales lead
to a change of 1.0$\%$ to $2.0\%$ and 8.0$\%$ in the signal $A \times \epsilon$ and cross section, respectively.  
In the case of the D1 (D5)  spin-independent operator, values of $M_*$ below 31~GeV and 5~GeV 
(585~GeV and 156~GeV)
are excluded at 90$\%$ CL for $\mchi$ equal to 1~GeV and 1.3~TeV, respectively.
Values of $M_*$ below 585~GeV and 100~GeV (794~GeV and 188~GeV) are
excluded for the D8 (D9) spin-dependent operator for $\mchi$ equal to 1~GeV and 1.3~TeV, respectively.
These results can be 
translated into upper limits on the nucleon-WIMP interaction cross section using the 
prescription in Refs.~\cite{wimps,fixed}.  Figure~\ref{fig:wimps} shows  
90$\%$ CL upper limits on the nucleon-WIMP cross section as a function of $\mchi$.  
In the case of the D1 (D5) spin-independent interaction, nucleon-WIMP cross 
sections above $2.7 \times 10^{-39}$~cm${}^{2}$  and 
$5.8 \times 10^{-34}$~cm${}^{2}$ ($2.2 \times 10^{-39}$~cm${}^{2}$  and 
$1.7 \times 10^{-36}$~cm${}^{2}$)  
are excluded at 90$\%$ CL for $\mchi = 1$~GeV and $\mchi = 1.3$~TeV, respectively.
Spin-dependent interactions cross sections in the range 
$7.6 \times 10^{-41}$~cm${}^{2}$  to $3.4 \times 10^{-37}$~cm${}^{2}$ 
($2.2 \times 10^{-41}$~cm${}^{2}$ to $2.7 \times 10^{-38}$~cm${}^{2}$) 
are excluded at 90$\%$ CL for the D8 (D9) operator  and  
$\mchi$ varying between 1~GeV and 1.3~TeV. 
The quoted observed limits on $M_*$ typically decrease by 2$\%$ to 10$\%$ if  the $-1\sigma$  
theoretical uncertainty 
is considered. This   
translates into a 10$\%$ to $50\%$ increase of the quoted nucleon-WIMP cross section limits.
The exclusion in the region  $ 1$~GeV $< \mchi < 3.5$~GeV 
($1$~GeV $ < \mchi < 1$~TeV) for spin-independent (spin-dependent) nucleon-WIMP interactions
is driven by the results from collider experiments with the assumption 
of the validity of the effective theory.
The  cross section upper limits  
improve upon  CDF results~\cite{cdf_wimps} 
and are similar to those obtained by 
the CMS experiment~\cite{cms} 
which uses 
axial-vector operators 
to describe 
spin-dependent interactions.

In summary, we report results on the search for new phenomena in 
events with an energetic photon and large missing transverse momentum 
in proton-proton collisions at $\sqrt{s}=7$~TeV at the LHC, 
based on ATLAS data corresponding to an integrated luminosity of 4.6~fb${}^{-1}$. 
The measurements are in agreement with the 
SM predictions for the background.  The results are translated into 
model-independent 90$\%$ and 95$\%$ confidence level upper limits
on $\sigma \times A \times \epsilon$ of
5.6~fb and 6.8~fb, respectively. The results are presented in terms
of improved limits on $M_D$ versus the number of extra spatial dimensions 
in  the ADD  model and upper limits on the spin-independent and spin-dependent
contributions to the 
nucleon-WIMP elastic cross section as a function of the WIMP mass.



We thank CERN for the very successful operation of the LHC, as well as the
support staff from our institutions without whom ATLAS could not be
operated efficiently.

We acknowledge the support of ANPCyT, Argentina; YerPhI, Armenia; ARC,
Australia; BMWF and FWF, Austria; ANAS, Azerbaijan; SSTC, Belarus; CNPq and FAPESP,
Brazil; NSERC, NRC and CFI, Canada; CERN; CONICYT, Chile; CAS, MOST and NSFC,
China; COLCIENCIAS, Colombia; MSMT CR, MPO CR and VSC CR, Czech Republic;
DNRF, DNSRC and Lundbeck Foundation, Denmark; EPLANET and ERC, European Union;
IN2P3-CNRS, CEA-DSM/IRFU, France; GNSF, Georgia; BMBF, DFG, HGF, MPG and AvH
Foundation, Germany; GSRT, Greece; ISF, MINERVA, GIF, DIP and Benoziyo Center,
Israel; INFN, Italy; MEXT and JSPS, Japan; CNRST, Morocco; FOM and NWO,
Netherlands; RCN, Norway; MNiSW, Poland; GRICES and FCT, Portugal; MERYS
(MECTS), Romania; MES of Russia and ROSATOM, Russian Federation; JINR; MSTD,
Serbia; MSSR, Slovakia; ARRS and MVZT, Slovenia; DST/NRF, South Africa;
MICINN, Spain; SRC and Wallenberg Foundation, Sweden; SER, SNSF and Cantons of
Bern and Geneva, Switzerland; NSC, Taiwan; TAEK, Turkey; STFC, the Royal
Society and Leverhulme Trust, United Kingdom; DOE and NSF, United States of
America.

The crucial computing support from all WLCG partners is acknowledged
gratefully, in particular from CERN and the ATLAS Tier-1 facilities at
TRIUMF (Canada), NDGF (Denmark, Norway, Sweden), CC-IN2P3 (France),
KIT/GridKA (Germany), INFN-CNAF (Italy), NL-T1 (Netherlands), PIC (Spain),
ASGC (Taiwan), RAL (UK) and BNL (USA) and in the Tier-2 facilities
worldwide.


\onecolumngrid
\clearpage 
\begin{flushleft}
{\Large The ATLAS Collaboration}

\bigskip

G.~Aad$^{\rm 48}$,
T.~Abajyan$^{\rm 21}$,
B.~Abbott$^{\rm 111}$,
J.~Abdallah$^{\rm 12}$,
S.~Abdel~Khalek$^{\rm 115}$,
A.A.~Abdelalim$^{\rm 49}$,
O.~Abdinov$^{\rm 11}$,
R.~Aben$^{\rm 105}$,
B.~Abi$^{\rm 112}$,
M.~Abolins$^{\rm 88}$
% ...
\bigskip

% ...
$^{11}$ Institute of Physics, Azerbaijan Academy of Sciences, Baku, Azerbaijan\\
$^{12}$ Institut de F\'{i}sica d'Altes Energies and Departament de F\'{i}sica de la Universitat Aut\`{o}noma de Barcelona and ICREA, Barcelona, Spain\\
% ...
$^{21}$ Physikalisches Institut, University of Bonn, Bonn, Germany\\
% ...
$^{48}$ Fakult\"{a}t f\"{u}r Mathematik und Physik, Albert-Ludwigs-Universit\"{a}t, Freiburg, Germany\\
$^{49}$ Section de Physique, Universit\'{e} de Gen\`{e}ve, Geneva, Switzerland\\
% ...
$^{88}$ Department of Physics and Astronomy, Michigan State University, East Lansing MI, United States of America\\
% ...
$^{105}$ Nikhef National Institute for Subatomic Physics and University of Amsterdam, Amsterdam, Netherlands\\
% ...
$^{111}$ Homer L. Dodge Department of Physics and Astronomy, University of Oklahoma, Norman OK, United States of America\\
$^{112}$ Department of Physics, Oklahoma State University, Stillwater OK, United States of America\\
% ...
$^{115}$ LAL, Universit\'{e} Paris-Sud and CNRS/IN2P3, Orsay, France\\
% ...
$^{s}$ Also at Manhattan College, New York NY, United States of America\\
% ...
\end{flushleft}



\begin{thebibliography}{99}
\bibitem{coord}   ATLAS uses a cylindrical coordinate system about the beam axis with
polar angle $\theta$ and azimuthal angle $\phi$. 
Anti-clockwise beam 
direction defines the positive $z$-axis, while the positive $x$-axis is defined as pointing from the collision 
point to the center of the LHC ring and the positive $y$-axis points upwards.  We define transverse energy 
$E_{\rm T} = E \ {\rm sin}  \theta$, 
transverse momentum $\pt = p \ {\rm sin} \theta$, and pseudorapidity $\eta = - {\rm ln}({\rm tan}(\theta/2))$.
\bibitem{masses} The masses of strange
and charm quarks are relevant for the cross sections of the D1 operator
and they are set to 0.1~GeV and
1.42~GeV, respectively.
\bibitem{fixed} In consultation with the authors of Ref.\cite{wimps}, a factor $4.7 \times
10^{-39} \rm{cm}^{2}$ is used in the cross section formula for D8 and D9 operators 
instead of the quoted $9.18 \times 10^{-40} {\rm cm}^{2}$.
\end{thebibliography}
\end{document}